\documentclass[conference]{IEEEtran}

\usepackage[left=1.62cm,right=1.62cm,top=1.9cm]{geometry}

\usepackage{mathrsfs}
\usepackage{cite}
\usepackage{graphicx}
\usepackage{amsmath,mathtools}
\usepackage{amssymb}
\usepackage{amsfonts}
\usepackage{tikzpagenodes}
\usepackage{lipsum}

\usepackage{xcolor}
\usepackage{amsthm}
\usepackage{bbm}
\begin{document}
%
\title{Achievable Rate Analysis in Molecular Channels with Reset-Counting Fully Absorbing Receivers}

%
%
%

\author{Fardad~Vakilipoor,
Luca~Barletta, Stefano~Bregni, and
Maurizio~Magarini
\\
Department of Electronic, Information and Bioengineering, Politecnico di Milano, Milan, Italy\\
\IEEEauthorblockA{
\{name.surname\}@polimi.it
}
}

\maketitle
\begin{tikzpicture}[remember picture,overlay]
    \node[align=center,text=gray] at ([yshift=1em]current page text area.north) {Submitted to IEEE Global Communications Conference, 4–8 December 2023, Kuala Lumpur, Malaysia};
  \end{tikzpicture}%

\begin{abstract}
In this paper, we investigate the achievable rate of a diffusive Molecular Communication (MC) channel with fully absorbing receiver, which counts particles absorbed along each symbol interval and resets the counter at every interval (reset-counting). The MC channel is affected by a memory effect and thus inter-symbol interference (ISI), due to the delayed arrival of molecules. To reduce complexity, our analysis is based on measuring the channel memory as an integer number of symbol intervals and on a single-sample memoryless detector. Thus, in our model the effect of released particles remains effective for a limited number of symbol intervals. We optimize the detector threshold for maximizing capacity, approximate as Gaussian the received signal distribution, and calculate the channel mutual information affected by ISI, in the case of binary concentration shift keying modulation. To the best of our knowledge, in literature there are no previous investigations on the achievable rate in this type of system. Our results demonstrate that, in general, the optimal input probability distribution achieving the maximum achievable rate may be not uniform. In particular, when the symbol interval is small (strong ISI), the maximum achievable rate does not occur with equiprobable transmission of bits.
\end{abstract}


\IEEEpeerreviewmaketitle

\section{Introduction}

Molecular Communication (MC) is an interdisciplinary communication paradigm, which relies on particle propagation as a mean of information transmission. MC has natural and artificial forms. Natural MC, which has evolved over millions of years, has great potential for investigating information exchange in biological systems. On the other hand, artificial MC is a human field that studies communication systems based on the principles of natural MC. One of the advantages of MC is its potential for use in environments where electromagnetic communication is not possible or desirable, such as in targeted drug delivery, nanomedicine and implantable devices, for which electromagnetic radiation can be harmful or interfere~\cite{chude2017molecular,cao2019diffusive}. 

Various aspects of MC systems have been studied, including active vs. passive receivers, instantaneous vs. continuous release of molecules, and for different boundary conditions of the physical channel~\cite{jamali2019channel}. In order to better understand this novel communication paradigm, an analysis from the information theory perspective would give new insights into MC and even improve the system performance in artificial counterparts. 
\subsection{Related Literature}
\par Channel capacity is a crucial metric to quantify communication system's ability to transmit information from the sender to the receiver~\cite{shannon1948mathematical}. Also in MC, the investigation of channel capacity is required, including memory effects causing inter-symbol interference (ISI), constrained energy, slow propagation, and unique statistical characterization~\cite{gohari2016information}. 

Early works on MC channel capacity considered transparent receivers, which do not interact with information particles (IP)~\cite{pierobon2012capacity}. 
However, in practice, receivers commonly bind with the IP through a reaction process in nature. 

The binding process can be related to the concentration of the hitting particles and equivalently to the number of particles absorbed by the receivers. Therefore, more recently, the evaluation of the capacity and bounds in diffusive MC channels with fully absorbing (FA) receivers was attempted. In~\cite{ratti2020upper}, upper and lower bounds of the channel capacity were evaluated, assuming Poisson and Gaussian models for the statistical characterization of the received signal. 

In ~\cite{ratti2021bounds}, the received signal was modeled as a Poisson random variable and bounds were determined for the constrained capacity of a diffusive MC system, which uses concentration shift keying (CSK) to modulate information. The lower bound was derived from the mutual information (MI), calculated as the difference between marginal entropy of the output and the conditional entropy of the output given the input. On the other hand, the upper bound was based on the definition of MI via the Kullback-Leibler divergence. 

In~\cite{liu2020channel}, the channel capacity for different reaction rates of the absorbing receiver was evaluated assuming uniform bit probability, although in capacity analysis the optimum input distribution in the transmission of bits would be expected. Moreover, this work considered the threshold of the memoryless detector as a predefined constant, which may not ensure optimal detection and maximum MI computation.
\subsection{Motivation and Contribution}

\par In this paper, we consider an FA receiver that, under perfect symbol synchronization assumption, counts particles absorbed along each symbol interval and resets the counter at the beginning of the next interval. The MC channel introduces a memory effect, due to the delayed arrival of molecules, and thus inter-symbol interference (ISI). 

In electronics, sample-and-hold circuits realize such resetting function. Brain neurotransmitters perform a similar task. Through reuptake mechanisms, the transporter proteins are responsible for removing neurotransmitters from the synaptic cleft, resetting their concentration and terminating their signaling effects~\cite{mukherjee2002regulation}. Observing that in nature and electronic circuits, we felt motivated to introduce such receiving mechanism and investigate the achievable rate, when the channel impulse response (CIR) varies with the transmission rate.

Unlike prior works, we estimate the channel memory length measured as an integer number of symbol intervals, to avoid the computational complexity arising from considering all possible permutations of previously transmitted bits. In our analysis, we adopt a single-sample memoryless detector, even though a multi-sample receiver may perform better due to the strong memory effect of the diffusive channel. Therefore, we refer to our capacity calculation as the memoryless capacity. 
Despite previous works that assumed a fixed threshold, we optimize the threshold to maximize the memoryless capacity. 

By characterizing the reset-counting receiver, we show that the achievable rate on the MC channel has a maximum for some optimal data transmission rate and binary input probability. We demonstrate that, in general, the optimal input distribution yielding the maximum
achievable rate may be not uniform. In particular, there is a value of symbol interval, over which the maximum is achieved with non equiprobable distribution. Note that shorter symbol interval implies higher ISI.
\subsection{Outline}
\par The paper is organized as follows. Section~\ref{sec:system_model} presents the system model, the memory calculation, and the CIR. Section~\ref{sec:capacity} explains the validity of the Gaussian approximation for channel modeling and formulates the memoryless channel capacity and achievable rate. Section~\ref{sec:simulation_results} presents numerical results of MI and achievable rate for different symbol intervals, input probabilities, and external noise power. Concluding remarks are provided in Section~\ref{sec:conclusion}.
\subsection{Notation}
\par Vectors are denoted as roman bold ($\mathbf{x}$), random variables as uppercase italic ($X$), their realizations as lowercase italic ($x$). The Hamming weight operator of a binary vector $\mathbf{x}\in\{0,1\}^n$ is denoted as $w_{H}(\mathbf{x})$ and counts the number of $1$'s in $\mathbf{x}$. The \textit{Q} function and complementary error function are defined as
\begin{equation}
    Q(z) = \frac{1}{2}\mathrm{erfc}\left ( \frac{z}{\sqrt{2}} \right )= \frac{1}{\sqrt{2\pi}}\int_{z}^{\infty} e^{-\frac{y^2}{2}}dy~.
\end{equation}

\section{System Model and Analysis}\label{sec:system_model}
This work considers a communication system made of a point transmitter, a diffusion-based channel and an FA spherical receiver. At the beginning of each symbol interval of duration~$T_{\mathrm{sym}}$ where "1" is sent, the transmitter releases a pulse of $N_{\mathrm{T}}$ IPs. The receiver counts the number of particles absorbed within each symbol interval and resets the counter at the beginning of the next interval. We believe that this mechanism is not far from reality~\cite{upadhyay2019emerging}. The IPs diffuse through the medium between transmitter and receiver with constant diffusion coefficient $D~[\mu \mathrm{m}^2/\mathrm{s}]$.

The receiver's absorption property stems from the reaction between receiver and IP. In effect, the counting process is tantamount to measuring the concentration of desired particles at the receiver, resulting from the interaction between its surface and particles. In a biological environment, enzymes can be secreted by the receiver to eliminate effects resulting from past reactions, thus enabling resetting~\cite{awan2017improving}. 

\par The propagation of diffusive particles is governed by Fick's second law, which relates the time derivative of the flux to the Laplacian of the concentration of molecules $c\left(d,t\right)$ at a given distance $d$ and time $t$, as
\begin{equation}
    \frac{\partial c \left( d,t \right)}{\partial t} = D \nabla^2c
    \left( d,t \right). \label{eq:2nd Fick}
\end{equation}
The initial and boundary conditions of~\eqref{eq:2nd Fick} vary depending on the MC system characteristics. Yilmaz~\textit{et al.}~\cite{yilmaz2014three} specified the boundary and initial conditions for an impulsive release of molecules, an unbounded environment, and an FA spherical receiver. They obtained the expression for the hitting rate of molecules onto the receiver surface, as a function of the distance~$d$ between the transmitter and the center of the receiver with radius $R$ at time~$t$. Then, assuming the independent random movement of the particles and the homogeneity of the medium, they derived the expected cumulative number of absorbed particles as 
 \begin{equation}
 N(t) = \frac{N_{\mathrm{T}}R}{d} \mathrm{erfc}\left ( \frac{d-R}{2\sqrt{Dt}} \right )~.\label{eq:CNAP}
 \end{equation}
\par In our study, we consider a binary concentration shift keying (BCSK) modulation, where IP release corresponds to "1" and no release corresponds to "0". At the receiver, the number of absorbed particles is counted and reset at the beginning of next interval. At the end of each symbol interval, the receiver returns a single sample, representing the total number of particles absorbed during that interval. Assuming that the receiver resets the counter right at the beginning of symbol intervals (\emph{i.e.} perfect synchronization between transmission and reset intervals at the receiver), we expect that the receiver observation changes by varying the duration of the symbol interval $T_{\mathrm{sym}}$, that is the inverse of the transmission rate. 

To compute the MI, we need to calculate the probability that particles hit the receiver. Since the total number of released particles is $N_{\mathrm{T}}$, if the counter has been not reset between the initial time of release until time $t$, the probability that a particle hits the receiver at time $t$ is $N(t)/N_{\mathrm{T}}$. If the counter is reset, instead, the probability that a particle released at $t=0$ hits the receiver within the $i$-th symbol interval is 
\begin{equation}
    p[i] = \frac{N(iT_{\mathrm{sym}})-N((i-1)T_{\mathrm{sym}})}{N_{\mathrm{T}}} \label{eq:pi}
\end{equation}
because a particle that has been absorbed at any time $t<(i$\,$-$\,$1) T_{\mathrm{sym}}$ does not have a second chance to hit the receiver.

When studying slow diffusive communication, it is important to quantify the effect of channel memory. To compute MI and transition probabilities between input and detected output, we need to account for all possible permutations of the preceding bit sequence. If channel memory spans $M$ symbols, there are $2^{M-1}$ different possible sequences to consider. The memory length depends on the transmission rate. In our model, it should be as smallest as possible, because evaluating $2^{M-1}$ permutations may make computation impractical. Moreover, due to the differential nature of~\eqref{eq:pi} and asymptotical convergence of~\eqref{eq:CNAP}, the probability of a particle being absorbed a long time after release eventually tends to 0. 

To get an estimate $M$ of the \emph{effective memory length} in terms of symbol intervals, being not unnecessarily long or so short to miss the effect of the released particles, we define
\begin{equation}
    M = \left\lceil \frac{T_{\alpha}}{T_{\mathrm{sym}}} \right\rceil 
\end{equation}
where $T_{\alpha}$ is the time required to reach some negligible hitting probability~$\alpha$, as given by
\begin{equation}
    \frac{R}{d}\left ( \mathrm{erfc}\left ( \frac{d-R}{2\sqrt{D(T_{\alpha}+T_{\mathrm{sym}}))}} \right ) -\mathrm{erfc}\left ( \frac{d-R}{2\sqrt{DT_{\alpha}}} \right ) \right )= \alpha~.\label{eq:alpha_criterion}
\end{equation}
Note that~\eqref{eq:alpha_criterion} is a transcendental equation with unknown~$T_{\alpha}$. We do not know an explicit solution for such equation. Hence, we solve it numerically by the \emph{regula-falsi} method. 

\begin{figure}
    \centering
    \includegraphics[width=0.9\columnwidth]{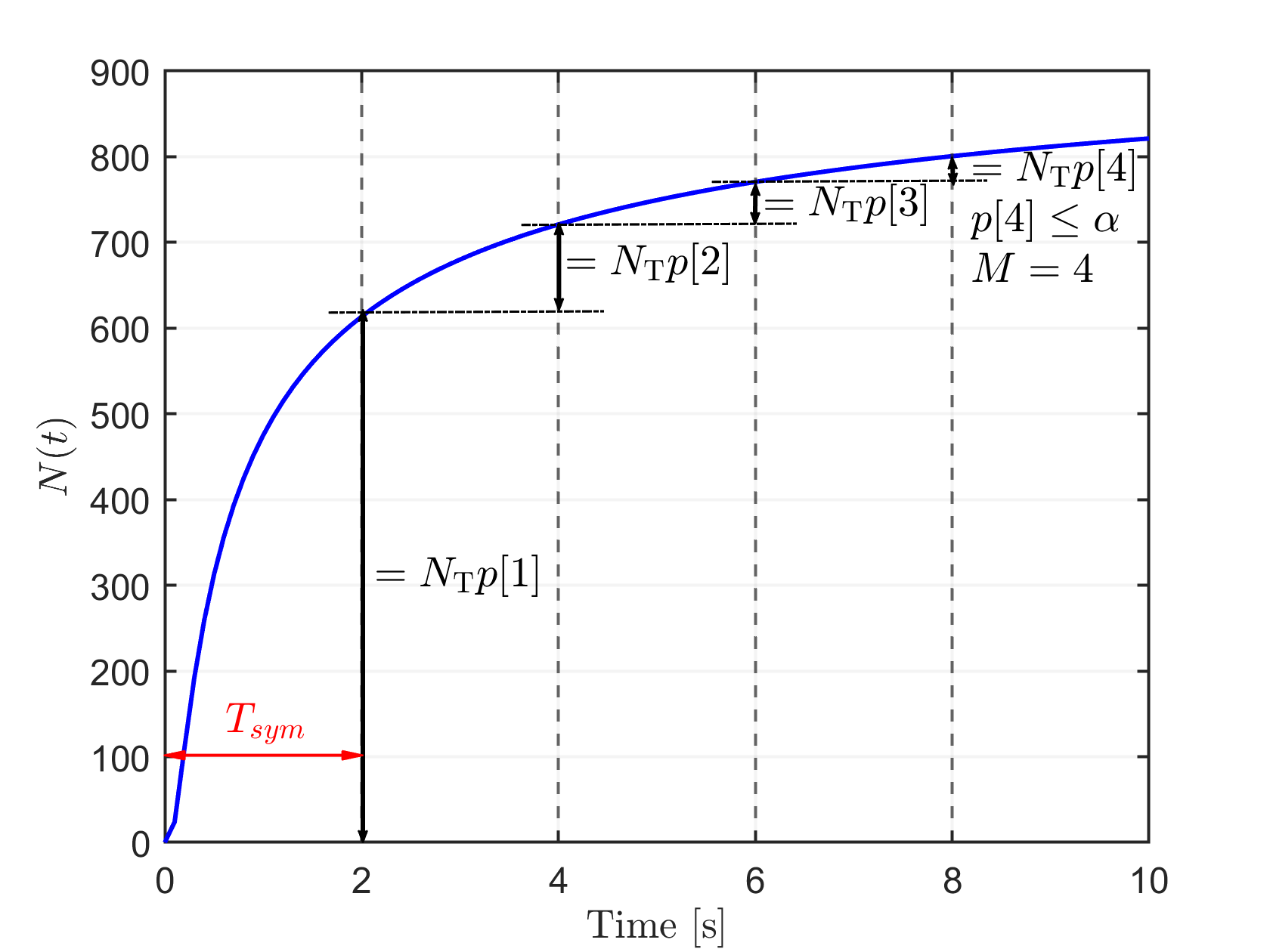}
    \caption{Expected cumulative number of absorbed particles over time without resetting (blue curve line) (system parameters as in Tab.~\ref{tab:param}, $T_{\mathrm{sym}}=2$~s, $M=4$).
   The expected number of absorbed particles within
each interval, resetting the counter, is  highlighted by vertical double arrows.}  
    \label{fig:CNAP}
\end{figure}

For example, Fig.~\ref{fig:CNAP} plots the expected cumulative number of absorbed particles over time without resetting the counter (blue curve line), computed by~\eqref{eq:CNAP} for a diffusive MC system modelled as above, with parameters set as in Tab.~\ref{tab:param}, and with symbol interval $T_{\mathrm{sym}}=2$~s. Here, the memory length results $M=4$. The expected number of absorbed particles within each symbol interval, resetting the counter at its beginning, is also highlighted as difference of values at interval boundaries.

\begin{figure*}[htp]

    \centering
    \includegraphics[width=.325\textwidth]{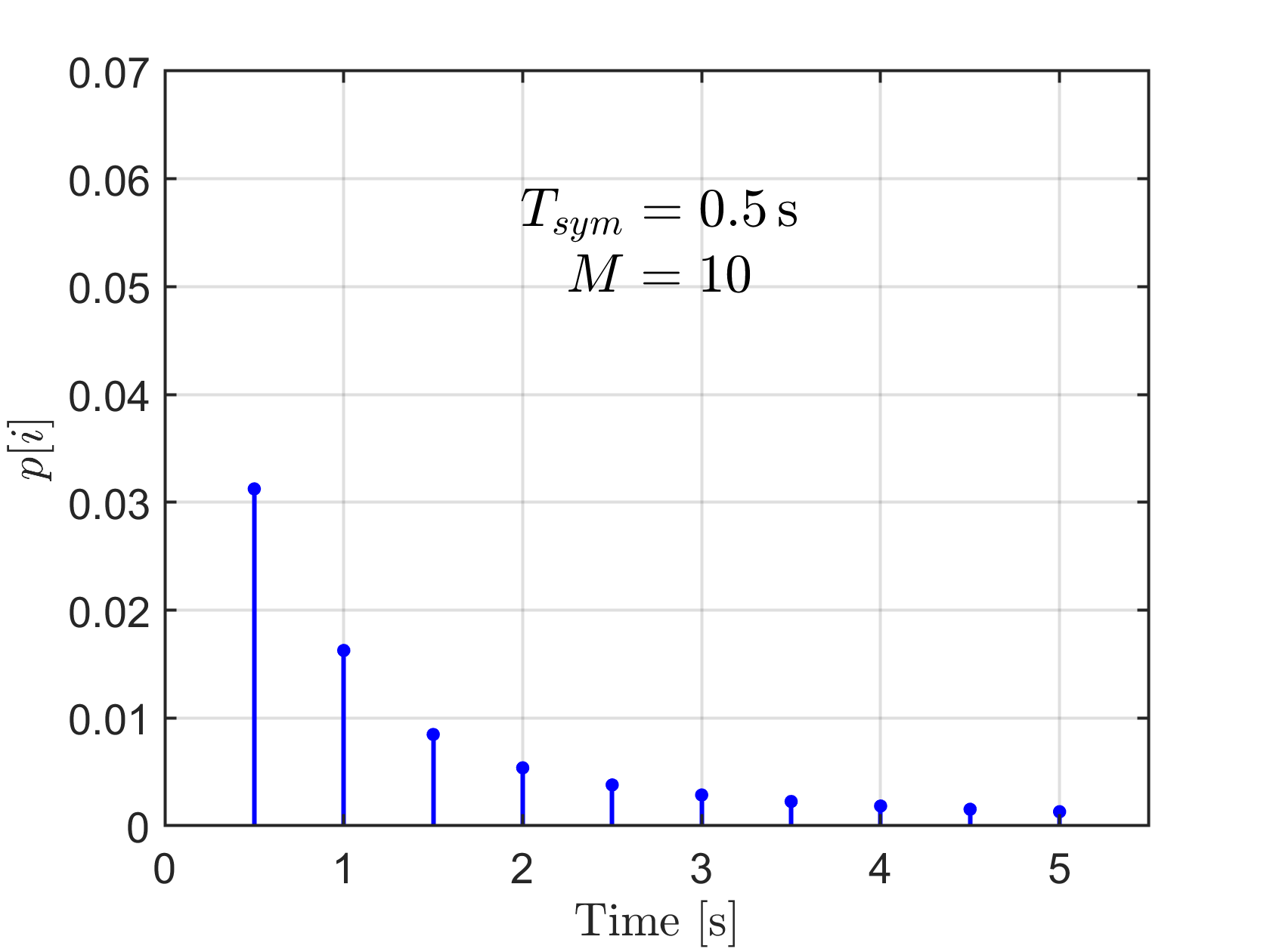}
    \includegraphics[width=.325\textwidth]{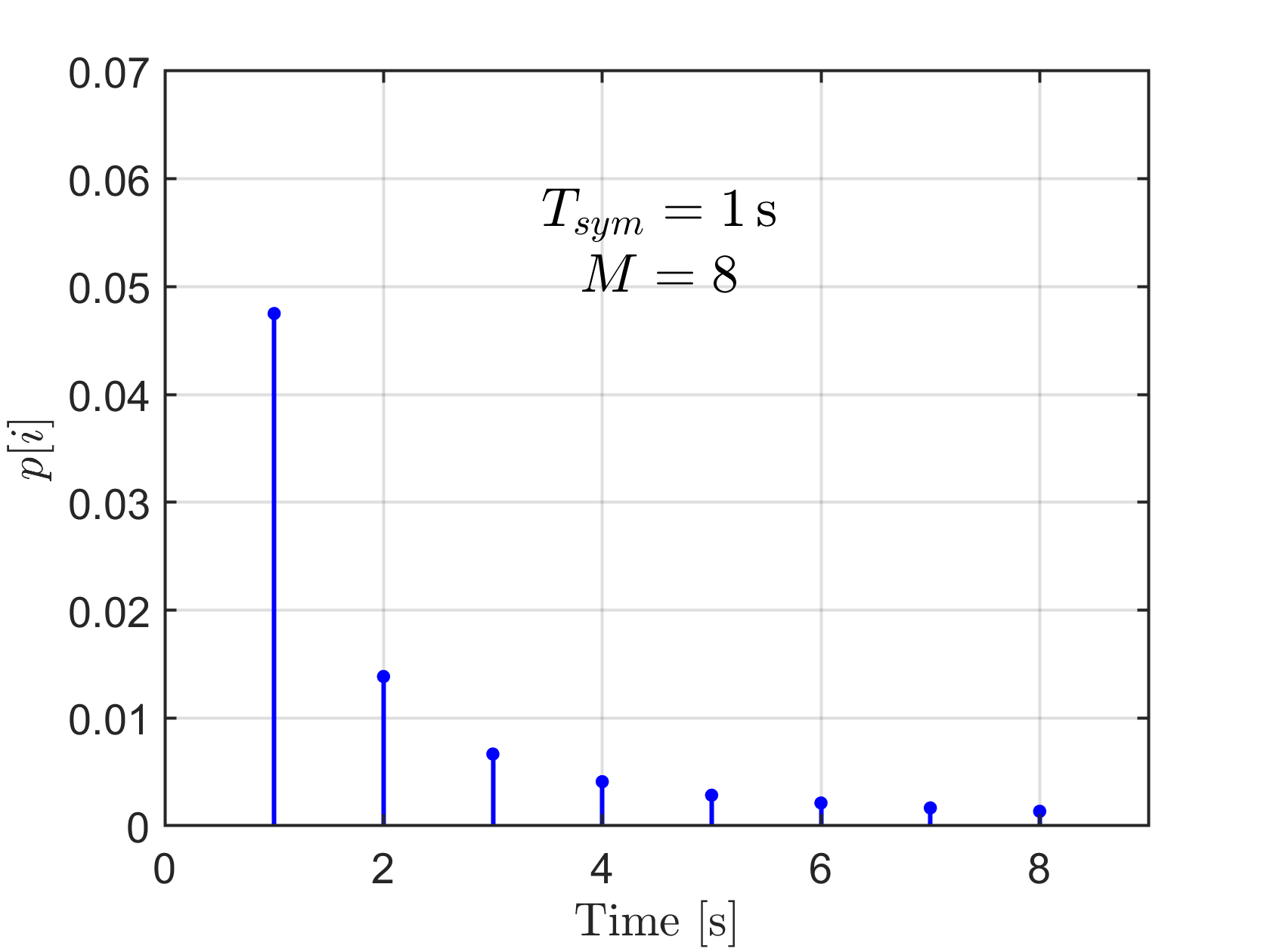}
    \includegraphics[width=.325\textwidth]{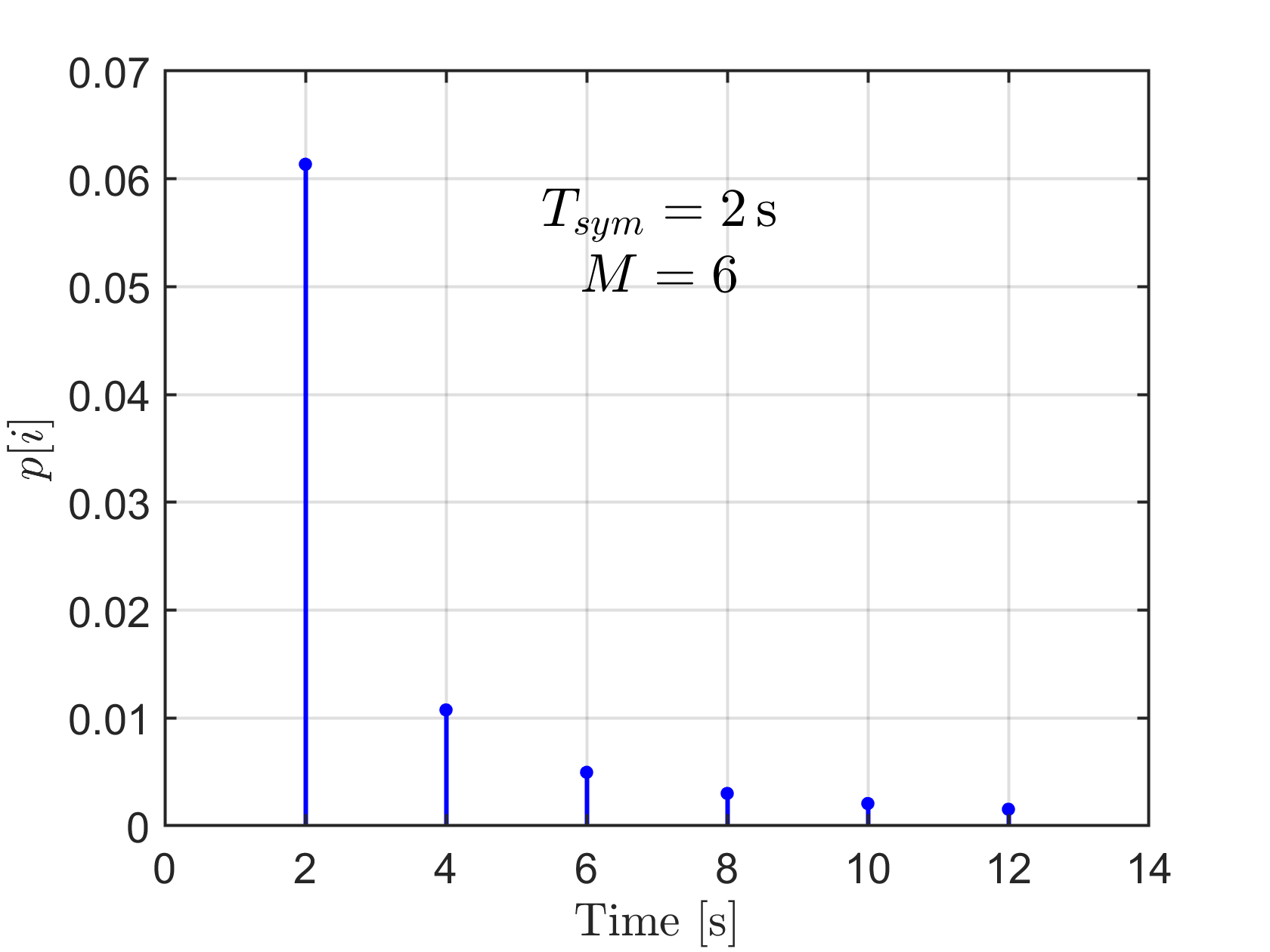}
    \caption{Distribution~\eqref{eq:pi} of the probability that particles are absorbed by a resetting receiver within the $i$-th interval for $T_{\mathrm{sym}}=0.5, 1, 2$~s. The channel memory length, resulting from \eqref{eq:alpha_criterion} for $\alpha=0.001$, increases with $T_{\mathrm{sym}}$ when measured in time units ($T_{\alpha}$), but decreases in terms of symbol intervals ($M$).}
    \label{fig:alpha_memory}
\end{figure*}

On the other hand, Fig.~\ref{fig:alpha_memory} plots the distribution~\eqref{eq:pi} of probability $p[i]$ that particles are absorbed by a resetting receiver within the $i$-th interval for different values of $T_{\mathrm{sym}}$. We observe that the memory length resulting from~\eqref{eq:alpha_criterion} increases with $T_{\mathrm{sym}}$ when measured in time units ($T_{\alpha}$), but decreases in terms of symbol intervals ($M$). The vector $\mathbf{p}=[p[1],\cdots,p[M]]$ represents the CIR of the system. We define $q[i]=1-p[i]$.

\section{Achievable Rate Analysis} \label{sec:capacity}
\par The received signal at the $i$-th symbol interval $N_{\mathrm{r}}[i]$ consists of the number of particles released for the $i$-th transmitted bit $N_{\mathrm{c}}[i]$ as well as of those released for previous bits and absorbed within the current interval $N_{\mathrm{p}}[i]$. We consider also an environment external noise $N_{\mathrm{ext}}$, due to random factors that increase or reduce the number of particles that the receiver counts in any interval. For example, a negative value of $N_{\mathrm{ext}}$ expresses the effect of extraneous molecules that unbind absorbed IPs. As obvious, $N_{\mathrm{ext}}$ is independent of $N_{\mathrm{c}}$ and $N_{\mathrm{p}}$. In conclusion, the observation at the $i$-th interval is the superposition of the current signal, of previously transmitted bits and of external noise, that is,
\begin{equation}
    N_{\mathrm{r}}[i] = N_{\mathrm{c}}[i] + N_{\mathrm{p}}[i] + N_{\mathrm{ext}}.
\end{equation}

Due to the nature of the absorption phenomenon, the success or failure in hitting the receiver by the particle can be seen as a random variable with Binomial distribution for $N_{\mathrm{T}}$ trials and success probability $\gamma$
\begin{equation}
    N\sim\mathcal{B}(N_{\mathrm{T}},\gamma).
\end{equation}
When the expected number of absorbed particles by the receivers is high, the Binomial distribution can be approximated by the Gaussian (Normal) distribution 
\begin{equation}
    N\sim\mathcal{N}(N_{\mathrm{T}}\gamma,N_{\mathrm{T}}\gamma(1-\gamma))~.
\end{equation}

In practice, to be this approximation valid, the probability that the Gaussian distribution generates negative values should be negligible. That is, the model parameters should be chosen to have mean $\mu$ and standard deviation $\sigma$ of the Gaussian distribution satisfying e.g.
$\mu>3\sigma$, which implies
\begin{equation}
    \frac{N_{\mathrm{T}}\gamma}{1-\gamma}>9~.\label{eq:valid_apx}
\end{equation}

The only scenario where this Gaussian approximation becomes weak is when the probability of particle received is extremely low. For example, in Fig.~\ref{fig:alpha_memory} we observe that a long time after transmission the probability of hitting gets very low. However, that happens only when the transmitter has not released any other particles for a long time. In conclusion, by considering a reasonable set of parameters that satisfies the condition above, in our study we can assume the Gaussian distribution as a valid approximation.
\par Let $s[i]\in\{0,1\}$ denote the transmitted bit associated to the $i$-th symbol interval and $g(\omega;\mu,\sigma^2)$ be a Gaussian probability density function (pdf) with mean $\mu$ and variance $\sigma^2$, where $g(\omega;0,0)=\delta(\omega)$ is the Dirac delta function. Then, the pdf of the current signal may be written as
\begin{equation}
    f_{N_\mathrm{c}[i]}(\omega) = \pi_{0}f_{N_\mathrm{c}[i]|S[i]=0}(\omega) + \pi_{1}f_{N_\mathrm{c}[i]|S[i]=1}(\omega)~,
\end{equation}
where $\pi_{1}=P_{S[i]}(1)$ and $\pi_{0}=1-\pi_{1}=P_{S[i]}(0)$ are the probability of transmitting "1" and "0", respectively. The conditional pdf of the current signal is then
\begin{equation}
    f_{N_\mathrm{c}[i]|S[i]=s[i]}(\omega) = g(\omega;s[i]N_{\mathrm{T}}p[1],s[i]N_{\mathrm{T}}p[1]q[1])~.
\end{equation}

Let the vector $\mathbf{s}\in\{0,1\}^{M-1}$ be a realization of $[s[i-1],\cdots,s[i-M+1]]$, that is the $M-1$ bits preceding the $i$-th interval. The conditional pdf of particles released in the past $M-1$ intervals and absorbed within the $i$-th  interval, given the sequence of preceding bits, is
\begin{align}
     \resizebox{.86\hsize}{!}{$\begin{aligned}
    f_{N_\mathrm{p}[i]|\mathbf{S}=\mathbf{s}}(\omega) = g\Bigg(&\omega;N_{\mathrm{T}}\sum^{M}_{j=2}s[i-j+1]p[j],\\ &N_{\mathrm{T}}\sum^{M}_{j=2}s[i-j+1]p[j]q[j]\Bigg)~.
     \end{aligned}$}
 \end{align}
Now, we can write the pdf of the received signal, due to the release of particles in the past, as
\begin{align}
     \resizebox{.85\hsize}{!}{$\begin{aligned}
    f_{N_\mathrm{p}[i]}(\omega) = \sum_{\forall{\mathbf{s}}}&\pi_{0}^{M-1-w_{H}(\mathbf{s})}\times \pi_{1}^{w_{H}(s)} f_{N_\mathrm{p}[i]|\mathbf{S}=\mathbf{s}}(\omega)~.
     \end{aligned}$}
 \end{align}
The external noise is assumed to follow a time-independent Gaussian distribution with pdf
\begin{equation}
    f_{N_{\mathrm{ext}}}(\omega) = g(\omega;\mu_{\mathrm{ext}},\sigma_{\mathrm{ext}}^{2}).
\end{equation}
We consider a memoryless binary detector with rule
\begin{equation}\label{eq:detect}
 \hat{S}[i]=\begin{cases}
    1 & \text{if $N_{r}[i]\geq\tau$}\\
    0 & \text{otherwise}.
  \end{cases}
\end{equation}
For each physical realization of the channel, we always look for the threshold $\tau$ that maximizes the MI.

To get the pdf of the received signals, two pdfs need to be considered first. The former is the pdf of the number of particles received in the $i$-th interval but released in previous intervals, including the environment noise, i.e. $N_{\mathrm{p}}[i]+N_{\mathrm{ext}}$. The latter is the pdf of particles received in the $i$-th interval, released in the current and previous intervals, including the external noise, that is $N_{\mathrm{c}}[i]+N_{\mathrm{p}}[i]+N_{\mathrm{ext}}$. As the pdf of the sum of random variables is given by the convolution of their pdfs, we get
\begin{align} 
    f_{N_{\mathrm{p}[i]}+N_{\mathrm{ext}}}(\omega) &= \sum_{\forall{\mathbf{s}}} \pi_{0}^{M-1-w_{H}(\mathbf{s})}\pi_{1}^{w_{H}(\mathbf{s})}\times \nonumber \\ &  g\Bigg(\omega;\mu_{ext}+N_{\mathrm{T}}\sum_{j=2}^{M}s[i-j+1]p[j],\nonumber\\ & \hspace{0.1cm}\sigma_{ext}^{2}+N_{\mathrm{T}}\sum^{M}_{j=2}s[i-j+1]p[j]q[j]\Bigg)
 \end{align}
\begin{align}
     \resizebox{.9\hsize}{!}{$\begin{aligned}
    f&_{N_\mathrm{c}[i]+N_\mathrm{p}[i]+N_{\mathrm{ext}}}(\omega) = \pi_{0}\sum_{\forall{\mathbf{s}}}\pi_{0}^{M-1-w_{H}(\mathbf{s})}\pi_{1}^{w_{H}(\mathbf{s})}\times \\ & \hspace{0.3cm}g\Bigg(\omega;\mu_{ext}+N_{\mathrm{T}}\sum^{M}_{j=2}s[i-j+1]p[j],\sigma_{ext}^{2}+N_{\mathrm{T}}\times \\&
    \hspace{0.3cm}\sum^{M}_{j=2}s[i-j+1]p[j]q[j]\Bigg) + \pi_{1}\sum_{\forall{\mathbf{s}}}\pi_{0}^{M-1-w_{H}(\mathbf{s})}\pi_{1}^{w_{H}(\mathbf{s})} \times\\& \hspace{0.3cm}g\Bigg(\omega;\mu_{ext}+N_{\mathrm{T}}\sum^{M}_{j=2}s[i-j+1]p[j]+ N_{\mathrm{T}}p[1],\\& \hspace{0.4cm}\sigma_{ext}^{2}+N_{\mathrm{T}}\sum^{M}_{j=2}s[i-j+1]p[j]q[j]+N_{\mathrm{T}}p[1]q[1]\Bigg)~.
     \end{aligned}$}
 \end{align}
Thus, the channel transition probabilities can be written as
\begin{align}
     \resizebox{.95\hsize}{!}{$\begin{aligned}
    P_{\hat{S}|S}(1|0) = \Pr(N_{\mathrm{p}}[i]&+N_{\mathrm{ext}} \geq\tau ) =  \sum_{\forall{\mathbf{s}}}\pi_{0}^{M-1-w_{H}(\mathbf{s})}\pi_{1}^{w_{H}(\mathbf{s})} \times \\& Q\Bigg(\frac{\tau-\mu_{ext}-N_{\mathrm{T}}\sum^{M}_{j=2}s[i-j+1]p[j]}{\sqrt{\sigma_{ext}^{2}+N_{\mathrm{T}}\sum^{M}_{j=2}s[i-j+1]p[j]q[j]}}\Bigg)
     \end{aligned}$}
 \end{align}
\begin{equation}
    P_{\hat{S}|S}(0|0) =\Pr(N_{\mathrm{p}}[i]+N_{\mathrm{ext}} <\tau ) = 1-\Pr(N_{\mathrm{p}}[i]+N_{\mathrm{ext}} \geq\tau )
\end{equation}
\begin{align}
     \resizebox{.98\hsize}{!}{$\begin{aligned}
    P_{\hat{S}|S}(1|1) =&\Pr(N_{\mathrm{c}}[i]+N_{\mathrm{p}}[i]+N_{\mathrm{ext}} \geq\tau) =\sum_{\forall{\mathbf{s}}}\pi_{0}^{M-1-w_{H}(\mathbf{s})}\pi_{1}^{w_{H}(\mathbf{s})} \times\\&Q\Bigg(\frac{\tau-\mu_{ext}-N_{\mathrm{T}}\sum^{M}_{j=2}s[i-j+1]p[j]-N_{\mathrm{T}}p[1]}{\sqrt{\sigma_{ext}^{2}+N_{\mathrm{T}}\sum^{M}_{j=2}s[i-j+1]p[j]q[j]+N_{\mathrm{T}}p[1]q[1]}}\Bigg)
     \end{aligned}$}
 \end{align}
\begin{align}
     \resizebox{1\hsize}{!}{$\begin{aligned}
   P_{\hat{S}|S}(0|1) = \Pr(N_{\mathrm{c}}[i]+N_{\mathrm{p}}[i]+N_{\mathrm{ext}} <\tau) = 1 - \Pr(N_{\mathrm{c}}[i]+N_{\mathrm{p}}[i]+N_{\mathrm{ext}} \geq\tau)~.
     \end{aligned}$}
 \end{align}

Given the transition probabilities, we can write the MI expression for the channel as follows
\begin{align}
     \resizebox{0.95\hsize}{!}{$\begin{aligned}
    I(S,\hat{S}) =& \pi_{0}P_{\hat{S}|S}(0|0)\log_{2}\frac{P_{\hat{S}|S}(0|0)}{\pi_{0}P_{\hat{S}|S}(0|0)+\pi_{1}P_{\hat{S}|S}(0|1)} +\\ 
    & \pi_{1}P_{\hat{S}|S}(0|1)\log_{2}\frac{P_{\hat{S}|S}(0|1)}{\pi_{0}P_{\hat{S}|S}(0|0)+\pi_{1}P_{\hat{S}|S}(0|1)}+\\
    & \pi_{0}P_{\hat{S}|S}(1|0)\log_{2}\frac{P_{\hat{S}|S}(1|0)}{\pi_{0}P_{\hat{S}|S}(1|0)+\pi_{1}P_{\hat{S}|S}(1|1)} +\\ 
    & \pi_{1}P_{\hat{S}|S}(1|1)\log_{2}\frac{P_{\hat{S}|S}(1|1)}{\pi_{0}P_{\hat{S}|S}(1|0)+\pi_{1}P_{\hat{S}|S}(1|1)}.\label{eq:I}
 \end{aligned}$}
 \end{align}

Accordingly, the \emph{memoryless channel capacity} is the maximum MI for any input probability
\begin{equation}
    C = \max_{\pi_{0}}\,I(S,\hat{S}).
\end{equation}
Finally, we define the \emph{achievable rate} [bit/s] as the MI divided by the symbol interval, that is $I(S,\hat{S})/T_{\mathrm{sym}}$. 
\section{Numerical Evaluation and Results}\label{sec:simulation_results}
We present a selection of results, which demonstrate that in general the optimal input distribution achieving the maximum achievable rate may be not uniform. Numerical evaluation was carried out with system parameters in Tab.~\ref{tab:param}, borrowed from~\cite{ferrari2022channel} except the external noise and~$\alpha$. We deliberately considered noise standard deviation and mean so that sometimes it becomes negative, which actually means that its effect is impeding an IP absorption. The parameter $\alpha$ is chosen such that the last sample of CIR is still valid according to \eqref{eq:valid_apx}.

\par
\begin{figure}
    \centering
    \includegraphics[width=0.9\linewidth]{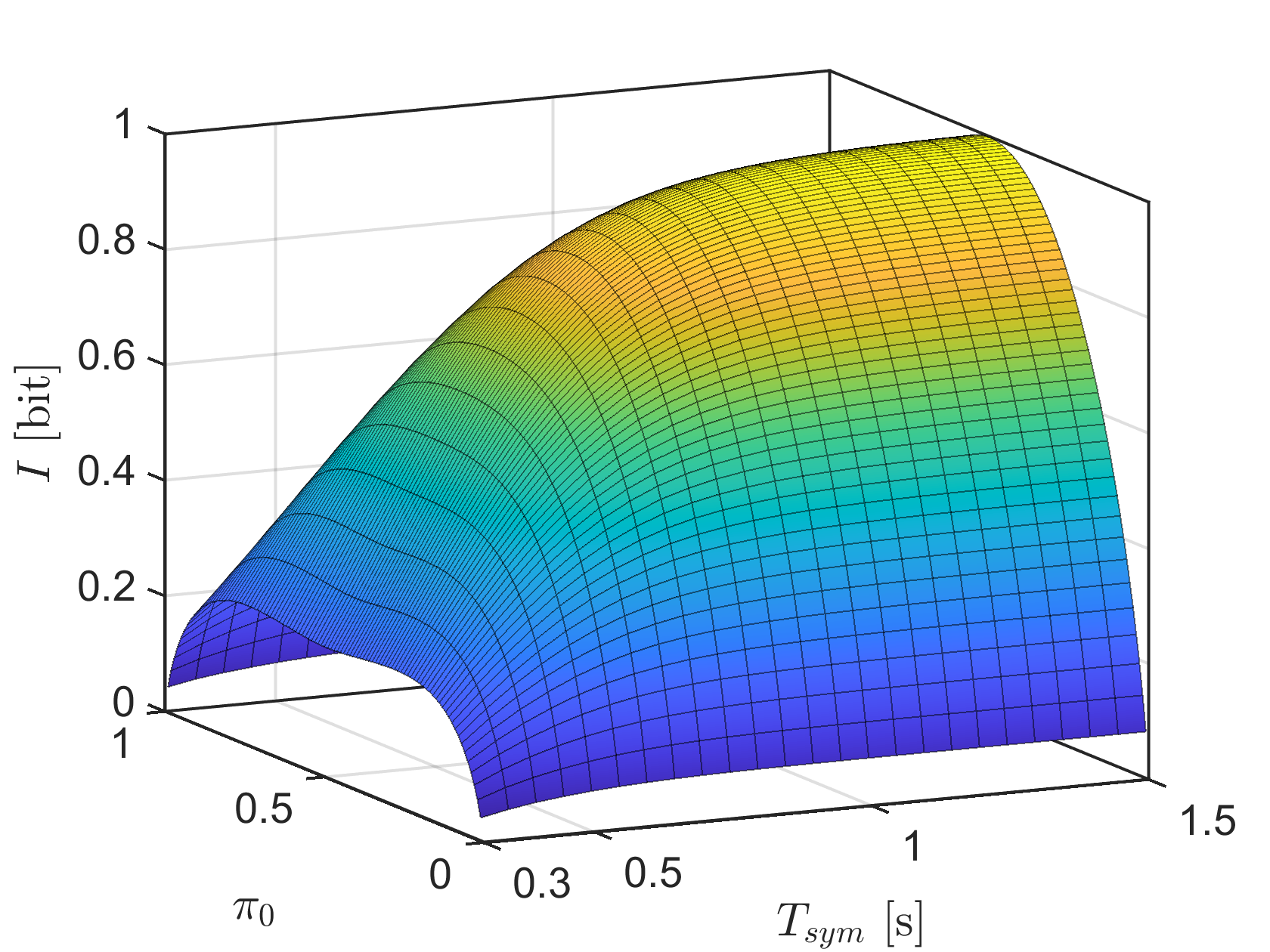}
    \caption{ MI as a function of the input distribution and symbol interval.}
    \label{fig:I}
\end{figure}

 Fig.~\ref{fig:I} plots the MI evaluated by \eqref{eq:I} varying the input distribution ($0 < \pi_{0} < 1$) and the symbol interval ($0.3$\,s$\leq T_{\mathrm{sym}} \leq 1.5$\,s). We observe that the parabolic shape of MI$(\pi_{0})$ in binary input transmission appears by  increasing $T_{\mathrm{sym}}$. The memoryless channel capacity is achieved for $T_{\mathrm{sym}}>0.5$\,s by an equiprobable input distribution $(\pi_{0}=0.5)$. 

\begin{figure}
    \centering
    \includegraphics[width=0.9\linewidth]{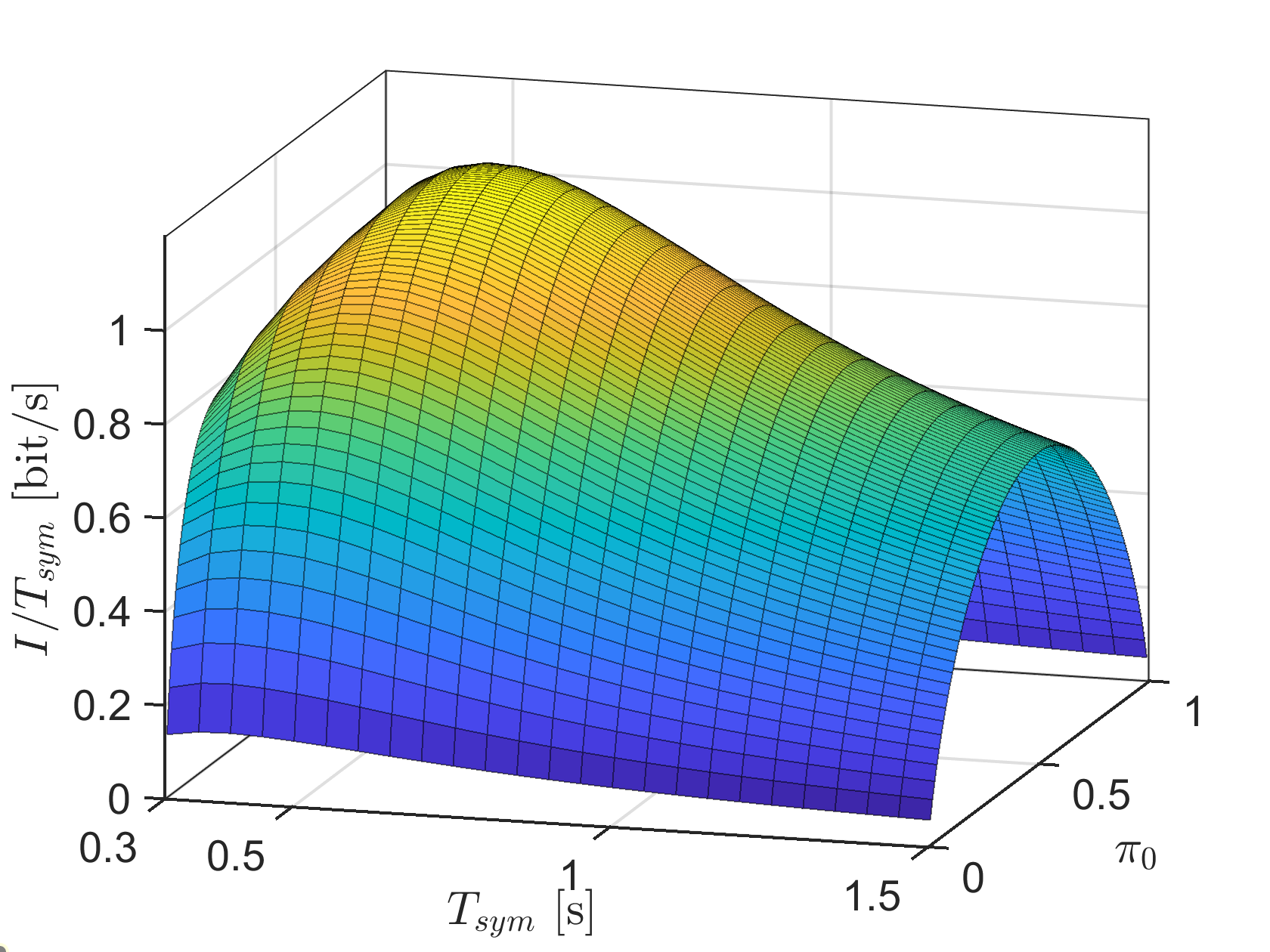}
    \caption{Achievable rate as a function of the input distribution and symbol interval.}
    \label{fig:AR_3D}
\end{figure}

The core finding of this work is demonstrated in Fig.~\ref{fig:AR_3D}, which plots the achievable rate for the same values of $\pi_{0}$ and $T_{\mathrm{sym}}$ as in Fig.~\ref{fig:I}. We observe that the achievable rate reaches its maximum $\approx\!1.18$ bit/s (corresponding to MI $\approx 0.71$\,bit) for $T_{\mathrm{sym}}\approx0.6$\,s and $\pi_{0}=0.5$. 

\begin{figure}
    \centering
    \includegraphics[width=0.9\linewidth]{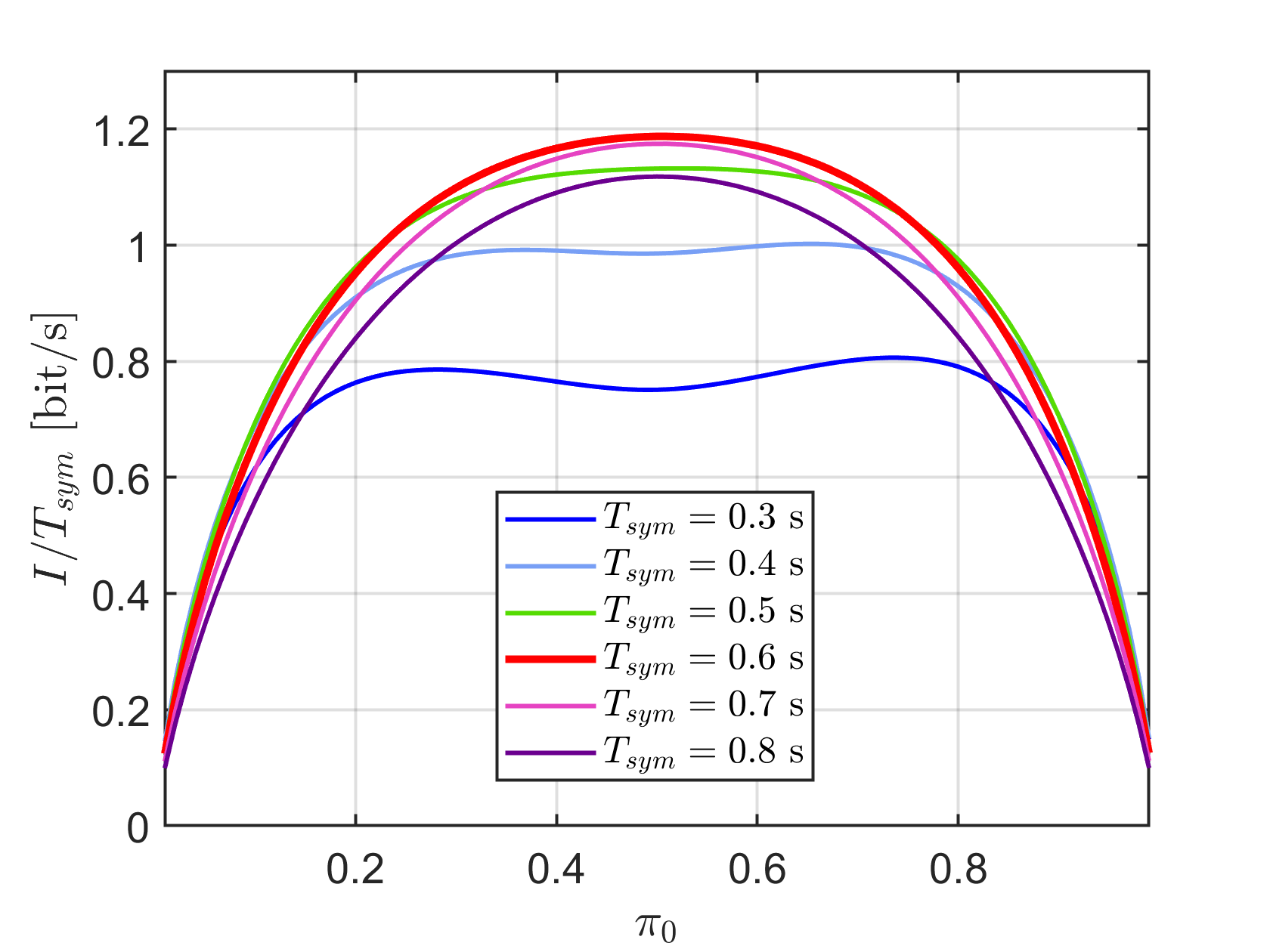}
    \caption{The achievable rate as a function of the input probability distribution.}
    \label{fig:AR_2D}
\end{figure}

It is worth noticing that the maximum achievable rate does not correspond to the maximum MI. Fig.~\ref{fig:AR_2D} plots the achievable rate as a function of $\pi_0$ evaluated for some values of $T_{\mathrm{sym}}$. Thus, these curves are vertical sections of the surface in Fig.~\ref{fig:AR_3D} for fixed $T_{\mathrm{sym}}$. Here, the first remark is that when the symbol interval $T_{\mathrm{sym}}$ is small, thus implying strong ISI, the maximum achievable rate does not occur with equiprobable transmission of bits ($\pi_1=\pi_0=0.5$).

The reason of the complex shape of the achievable rate curves in Fig.~\ref{fig:AR_2D} is not trivial: each $\pi_0$ is associated to a different channel,  which depends on the specific optimum detector threshold~$\tau$. The curve for $T_{\mathrm{sym}}=0.3$\,s exhibits two local maxima. The maximum at $\pi_{0}\approx0.28$ suggests transmitting fewer "0"s, which may seem counter-intuitive given the higher ISI associated with faster transmission rates. However, the other maximum at $\pi_{0}\approx0.75$ suggests transmitting more "0"s. This observation is sensible because the ISI increases with the transmission rate. By transmitting "1" less frequently, the ISI is reduced, yielding an improvement in the achievable rate. As expected, the maximum associated with $\pi_{0}\approx0.75$ is higher than the one associated with~$\pi_{0}\approx0.28$. 

\begin{figure}[t!]
    \centering
    \includegraphics[width=0.9\linewidth]{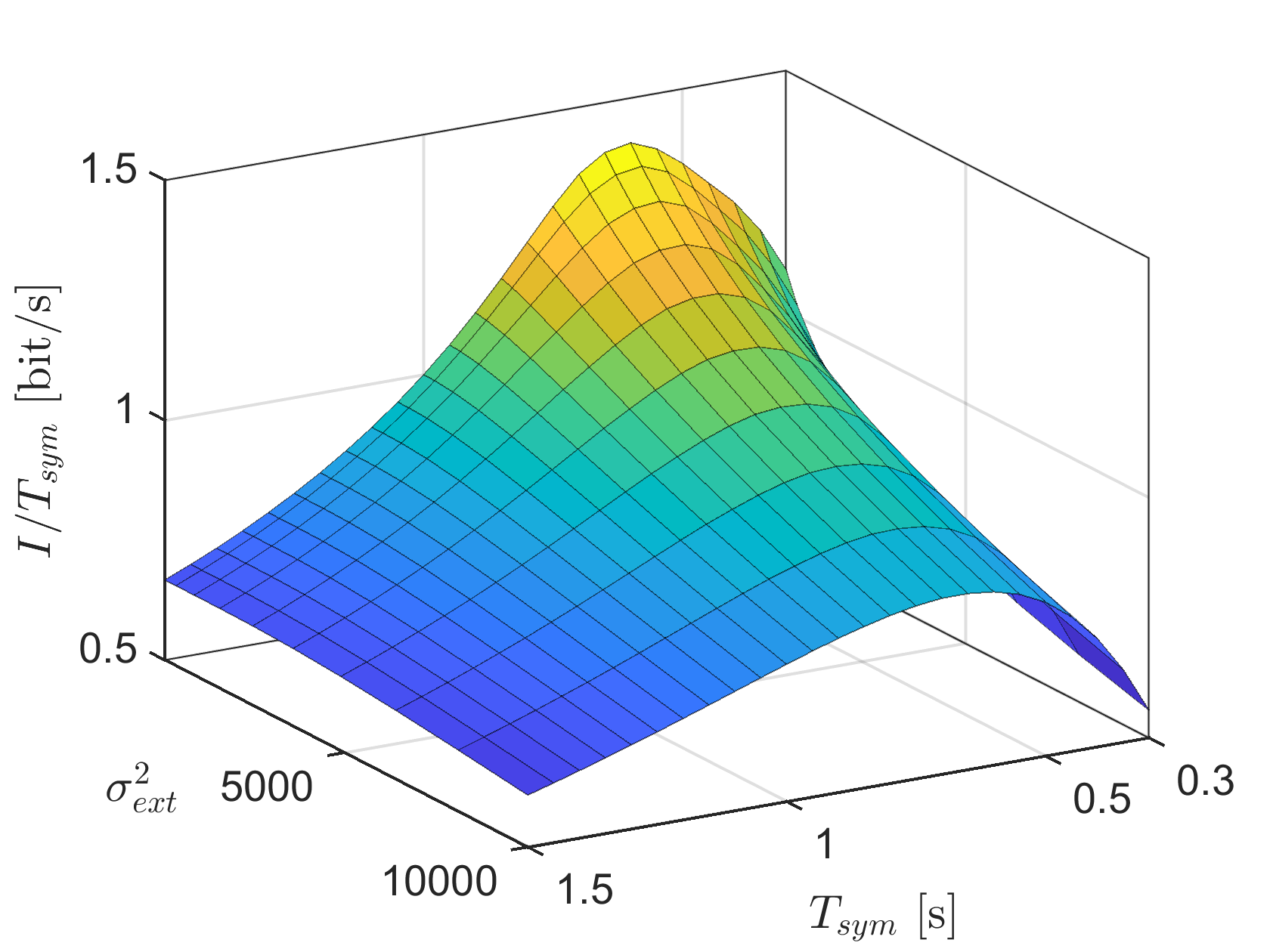}
    \caption{Achievable rate as a function of the external noise power and symbol interval with optimal input distribution.}
    \label{fig:AR_sigma_3D}
\end{figure}

Finally, Fig.~\ref{fig:AR_sigma_3D} plots the achievable rate, evaluated with optimum input distribution, for various values of external noise power and of symbol interval. As expected, we note that the achievable rate drops by increasing the noise power.
\begin{table}[!t]
\begin{center}
\caption{System parameters}
\label{tab:param}
\resizebox{0.48\textwidth}{!}{
 \begin{tabular}{|| c | c | c ||}
 \hline
 Variable & Definition & Value \\ [0.5ex] 
 \hline\hline
  $N_{\mathrm{T}}$ & Number of released molecules & $10^{4}$ \\ 
  \hline
 $R$ & Radius of the receiver $\mathcal{R}_i$ & $1$ $\mu$m\\ 
 \hline
 $d$ & Distance between transmitter and center of receiver & $10$ $\mu$m \\
 \hline
 $\alpha$ & Minimum acceptable probability & $0.001$ \\
 \hline
   $\mu_{ext}$ & Mean of the external noise signal & $50$ \\
 \hline
   $\sigma_{ext}$ & Standard deviation of the external noise signal & $50$ \\
 \hline
 $D$ & Diffusion coefficient for the signaling molecule & $79.4$ $\mu\text{m}^2/\text{s}$ \\
 \hline
\end{tabular}}
\end{center}
\end{table}
\section{Conclusions}\label{sec:conclusion}

We have investigated the achievable rate of a diffusive MC channel with FA receiver, which counts particles absorbed along each symbol interval and resets the counter at every interval. The MC channel is affected by memory and thus ISI, due to the delayed arrival of molecules. 
To reduce complexity, we have measured the effective memory length as an integer number of symbol intervals and considered a single-sample memoryless detector. Unlike previous works, we have also optimized the detector threshold to maximize capacity. We have approximated as Gaussian the received signal distribution and calculated the channel mutual information affected by ISI.

Our selection of numerical results demonstrate that, in general, the optimal input probability distribution achieving the maximum achievable rate may be not uniform. In particular, when the symbol interval $T_{\mathrm{sym}}$ is small, thus implying strong ISI, the maximum achievable rate does not occur with equiprobable transmission of bits ($\pi_{1}=\pi_{0}=0.5$).

\ifCLASSOPTIONcaptionsoff
  \newpage
\fi

\bibliographystyle{IEEEtran}
\bibliography{IEEEabrv,pulseshape}

\end{document}